\documentclass[10pt,twocolumn,letterpaper]{article}

\usepackage[pagenumbers]{cvpr}      

\usepackage{graphicx}
\usepackage{amsmath}
\usepackage{amssymb}
\usepackage{booktabs}
\usepackage{multirow}
\usepackage{enumitem}

\usepackage[table,xcdraw]{xcolor}
\usepackage{mathtools}
\usepackage{adjustbox}
\usepackage{indentfirst}
\usepackage{multirow}
\usepackage[T1]{fontenc}
\usepackage{mathrsfs}
\usepackage{array}
\usepackage{colortbl}
\usepackage{bm}
\usepackage{lipsum}
\usepackage{color}

\newcommand{\tabincell}[2]{\begin{tabular}{@{}#1@{}}#2\end{tabular}}  

\usepackage[pagebackref,breaklinks,colorlinks]{hyperref}

\usepackage[capitalize]{cleveref}
\crefname{section}{Sec.}{Secs.}
\Crefname{section}{Section}{Sections}
\Crefname{table}{Table}{Tables}
\crefname{table}{Tab.}{Tabs.}

\def\cvprPaperID{5086} 
\def\confName{CVPR}
\def\confYear{2022}

\begin{document}

\title{Deep Constrained Least Squares for Blind Image Super-Resolution}

\author{Ziwei Luo\,$^1$\quad
Haibin Huang\,$^2$\quad
Lei Yu\,$^1$\quad
Youwei Li\,$^1$\quad
Haoqiang Fan\,$^1$\quad
Shuaicheng Liu\,$^{3,1}$\thanks{Corresponding author.}  \\
$^1$\,Megvii Technology\quad$^2$\,Kuaishou Technology\\
$^3$\,University of Electronic Science and Technology of China\\
{\normalsize \url{https://github.com/megvii-research/DCLS-SR}}


}
\maketitle

\begin{abstract}


In this paper, we tackle the problem of blind image super-resolution(SR) with a reformulated degradation model and two novel modules. Following the common practices of blind SR, our method proposes to improve both the kernel estimation as well as the kernel based high resolution image restoration. To be more specific, we first reformulate the degradation model such that the deblurring kernel estimation can be transferred into the low resolution space. On top of this, we introduce a dynamic deep linear filter module. Instead of learning a fixed kernel for all images, it can adaptively generate deblurring kernel weights conditional on the input and yields more robust kernel estimation. Subsequently, a deep constrained least square filtering module is applied to generate clean features based on the reformulation and estimated kernel. The deblurred feature and the low input image feature are then fed into a dual-path structured SR network and restore the final high resolution result. To evaluate our method, we further conduct evaluations on several benchmarks, including Gaussian8 and DIV2KRK. Our experiments demonstrate that the proposed method achieves better accuracy and visual improvements against state-of-the-art methods. 

\end{abstract}


\section{Introduction}
In this work, we study the problem of image super-resolution,i.e., restoring high-resolution images from low-resolution inputs. Specially, we aim for single image super-resolution (SISR), where only one observation is given which is a more practical setting and with a wide range of downstream applications ~\cite{dong2014learning,kim2016accurate,lim2017enhanced,ledig2017photo,zhang2018image,wang2018esrgan,fritsche2019frequency,li2019feedback,zhang2019deep,haris2018deep}.   

Most existing works based on the classical SISR degradation model assuming that the input LR image ${\bf y}$ is a blurred and down-scaled HR image ${\bf x}$ with additional white Gaussian noise {\bf n}, given by
\begin{equation}
\centering
    {\bf y} = ({\bf x} \ast {\bf k}_h)_{\downarrow_s} + {\bf n},
\label{eq:class_degrad}
\end{equation}
where ${\bf k}_h$ is the blur kernel applied on $\bf x$, $\ast$ denotes convolution operation and $\downarrow_s$ denotes downsampling with scale factor $s$. 
Previous blind SR approaches ~\cite{gu2019blind,luo2020unfolding} generally solve this problem with a two-stage framework: kernel estimation from LR image and kernel based HR image restoration. 

We argue that although such a pipeline demonstrates reasonable performance for SR problem, there are two main drawbacks: First of all, it is difficult to accurately estimate blur kernels of HR space directly from LR images due to the ambiguity produced by undersampling step ~\cite{vandewalle2006frequency,park2003super}. And the mismatch between the estimated kernel and the real one will cause significant performance drop and even lead to unpleasant artifacts~\cite{zhang2018learning,gu2019blind,bell2019blind,hussein2020correction}.
Secondly, it is also challenging to find a suitable way to fully utilize the information of the estimated HR space kernel and LR space image. A common solution is to employ a kernel stretching strategy ~\cite{zhang2018learning,gu2019blind,luo2020unfolding}, where the principal components of the vectorized kernel are preserved and stretched into degradation maps with the same size as the LR input. These degradation maps then can be concatenated with the input image or its features to generate a clean HR image. However, the spatial relation of the kernel is destroyed by the process of vectorizing and PCA (Principal Component Analysis), which causes insufficient usage of the kernel.  The subsequent reconstruction network requires a huge effort to harmonize the inconsistent information between LR features and HR-specific kernels, limiting its performance in super-resolving images.

\begin{figure}[t]
\setlength{\abovecaptionskip}{-0.05in}
\setlength{\belowcaptionskip}{-0.1in}
\begin{center}
\includegraphics[width=.99\linewidth]{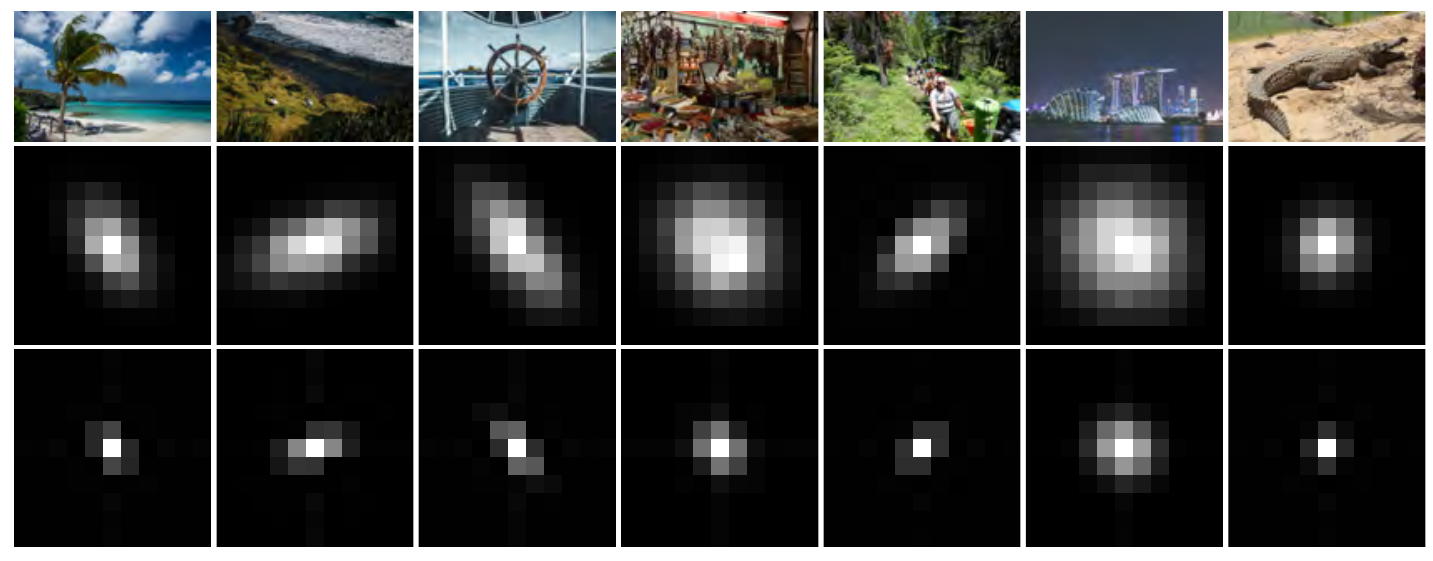}
\end{center}
\caption{Kernel reformulation examples. The top row and middle row are the LR images and the corresponding primitive kernels. The bottom row is the reformulated kernels.}
\label{fig:reform_kernels}
\end{figure}

Towards this end, we present a modified learning strategy to tackle the blind SR problem, which can naturally avoid the above mentioned drawbacks. Specifically, we first reformulate the degradation model in a way such that the blur kernel estimation and image upsampling can be disentangled. In particular, as shown in Fig.~\ref{fig:reform_kernels}, we derive a new kernel from the primitive kernel ${\bf k}_h$ and LR image. It transfers the kernel estimation into the LR space and the new kernel can be estimated without aliasing ambiguity. Based on the new degradation, we further introduce the dynamic deep linear kernel (DDLK) to provide more equivalent choices of possible optimal solutions for the kernel to accelerate training. 
Subsequently, a novel deep constrained least squares (DCLS) deconvolution module is applied in the feature domain to obtain deblurred features. DCLS is robust to noise and can provide a theoretical and principled guidance to obtain clean images/features from blurred inputs. Moreover, it dosen't require kernel stretching strategy and thus preserves the kernel's spatial relation information. Then the deblurred features are fed into an upsampling module to restore the clean HR images. As illustrated in Fig.~\ref{fig:ts}, the overall method has turned out to be surprisingly effective in recovering sharp and clean SR images.


The main contributions are summarized as follows:

\begin{itemize}
    \item We introduce a new practical degradation model derived from \cref{eq:class_degrad}. Such degradation maintains consistency with the classical model and allows us reliably estimate blur kernel from low-resolution space.
    
    \item We propose to use a dynamic deep linear kernel instead of a single layer kernel, which provides more equivalent choices of the optimal solution of the kernel, which is easier to learn.
    
    \item We propose a novel deconvolution module named DCLS that is applied on the features as channel-wise deblurring so that we can obtain a clean HR image.
    
    \item Extensive experiments on various degradation kernels demonstrate that our method leads to state-of-the-art performance in blind SR problems.
\end{itemize}

\vspace{-0.1in}

\section{Related work}

\noindent \textbf{Non-blind SR}
Since pioneering work SRCNN~\cite{dong2014learning} proposes to learn image SR with a three-layer convolution network, most subsequent works have focused on optimizing the network architectures~\cite{kim2016accurate,lim2017enhanced,shi2016real,zhang2018residual,zhang2018image,lai2017deep,zhang2017beyond,haris2018deep,kim2016deeply,tai2017image,zhang2020residual,luo2021ebsr,dai2019second} and loss functions~\cite{ledig2017photo,wang2018esrgan,johnson2016perceptual,yu2016ultra,lugmayr2020srflow,zhang2018unreasonable,wang2018recovering}. These CNN-based methods have achieved impressive performance on SISR with a predefined single degradation setting (e.g., bicubic downsampling). However, they may suffer significant performance drops when the predefined degradation kernel is different from the real one. 

Some non-blind SR approaches address the multiple degradation problem by restoring HR images with given the corresponding kernels. Specifically, SRMD~\cite{zhang2018learning} is the first method that concatenates LR image with a stretched blur kernel as inputs to obtain a super-resolved image under different degradations. Later, Zhang \textit{et al.}~\cite{zhang2019deep,zhang2020deep} incorporate advanced deblurring algorithms and extend the degradation to arbitrary blur kernels. UDVD~\cite{xu2020unified} improves the performance by incorporating dynamic convolution. Hussein \textit{et al.}~\cite{hussein2020correction} introduce a correction filter that transfers blurry LR images to match the bicubicly designed SR model. Besides, zero-shot methods~\cite{xu2020unified,soh2020meta} have also been investigated in non-blind SR with multiple degradations.  

\noindent \textbf{Blind SR}
Under the blind SR setting, HR image is recovered from the LR image degraded with unknown kernel~\cite{levin2009understanding,levin2011efficient,michaeli2013nonparametric}. Most approaches solve this problem with a two stage framework: kernel estimation and kernel-based HR image restoration. For the former,  KernelGAN~\cite{bell2019blind} estimates the degradation kernel by utilizing an internal generative adversarial network(GAN) on a single image, and applies that kernel to a non-blind SR approach such as ZSSR to get the SR result. Liang \textit{et al.}~\cite{liang2021flow} improve the kernel estimating performance by introducing a flow-based prior. Furthermore, Tao \textit{et al.}~\cite{tao2021spectrum} propose a spectrum-to-kernel network and demonstrate that estimating blur kernel in the frequency domain is more conducive than in spatial domain. For the latter, Gu \textit{et al.}~\cite{gu2019blind} propose to apply spatial feature transform (SFT) and iterative kernel correction (IKC) strategy for accurate kernel estimation and SR refinement. Luo \textit{et al.}~\cite{luo2020unfolding} develop an end-to-end training deep alternating network (DAN) by estimating reduced kernel and restoring HR image iteratively. However, both IKC and DAN are time-consuming and computationally costly. The modified version of DAN~\cite{luo2021endtoend} conducts a dual-path conditional block (DPCB) and supervises the estimator on the complete blur kernel to further improve the performance.

\section{Method}
We now formally introduce our method which consists of three main components given a reformation of degradation: A dynamic deep linear kernel estimation module and a deep constrained least squares module for kernel estimation and LR space feature based deblur. A dual-path network is followed to generate the clean HR output. We will first derive the reformulation and then detail each module.

\subsection{Degradation Model Reformulation}

Ideally, the blur kernel to be estimated and its corresponding image should be in the same low-resolution space such that the degradation can be transformed to the deblurring problem followed by a SISR problem with bicubic degradation~\cite{zhang2018learning,zhang2019deep}.
Towards this end, we propose to reformulate Eq.~(\ref{eq:class_degrad}) as
\begin{align}
\centering
    {\bf y} &= {\cal F}^{-1}\left({\cal F}\left(({\bf x} \ast {\bf k}_h)_{\downarrow_s}\right)\right) + {\bf n}\\
            &= {\cal F}^{-1}\left({\cal F}\left({\bf x}_{\downarrow_s}\right)\frac{{\cal F}\left(({\bf x} \ast {\bf k}_h)_{\downarrow_s}\right)}{{\cal F}\left({\bf x}_{\downarrow_s}\right)} \right) + {\bf n} \\
            &= {\bf x}_{\downarrow_s} \ast {\cal F}^{-1}\left(\frac{{\cal F}\left(({\bf x} \ast {\bf k}_h)_{\downarrow_s}\right)}{{\cal F}\left({\bf x}_{\downarrow_s}\right)} \right) + {\bf n},
\label{eq:derive_new_degrad}
\end{align}
where $\cal F$ denotes the Discrete Fourier Transform and ${\cal F}^{-1}$ denotes its inverse. Then let
\begin{equation}
\centering
    {\bf k}_l = {\cal F}^{-1}\left(\frac{{\cal F}\left(({\bf x} \ast {\bf k}_h)_{\downarrow_s}\right)}{{\cal F}\left({\bf x}_{\downarrow_s}\right)}\right),
\label{eq:derive_k}
\end{equation}
we can obtain another form of degradation:
\begin{equation}
\centering
    {\bf y} = {\bf x}_{\downarrow_s} \ast {\bf k}_l + {\bf n}.
\label{eq:new_degrad}
\end{equation}
In the \cref{eq:new_degrad}, ${\bf k}_l$ is derived from the corresponding ${\bf k}_h$ and applied on the downsampled HR image ${\bf x}_{\downarrow_s}$. 
To ensure numerical stability, we rewrite Eq.~(\ref{eq:derive_k}) with a small regularization parameter $\epsilon$:
\begin{equation}
\centering
    {\bf k}_l = {\cal F}^{-1}\left(\frac{\overline{{\cal F}({\bf x}_{\downarrow_s})}}{\overline{{\cal F}({\bf x}_{\downarrow_s})}{\cal F}({\bf x}_{\downarrow_s}) + \epsilon} {\cal F}\left(({\bf x} \ast {\bf k}_h)_{\downarrow_s}\right)\right),
\label{eq:derive_stable_k}
\end{equation}
where $\overline{{\cal F}(\cdot)}$ is the complex conjugate of ${\cal F}$. Fig.~\ref{fig:reform_kernels} illustrates the results of reformulating kernels by \cref{eq:derive_stable_k}. Based on the new degradation process, our goal is to estimate the blur kernel ${\bf k}_l$ and then restore HR image $\bf x$.


\begin{figure}[t]
\setlength{\abovecaptionskip}{-0.in}
\setlength{\belowcaptionskip}{-0.1in}
\begin{center}
\includegraphics[width=.9\linewidth]{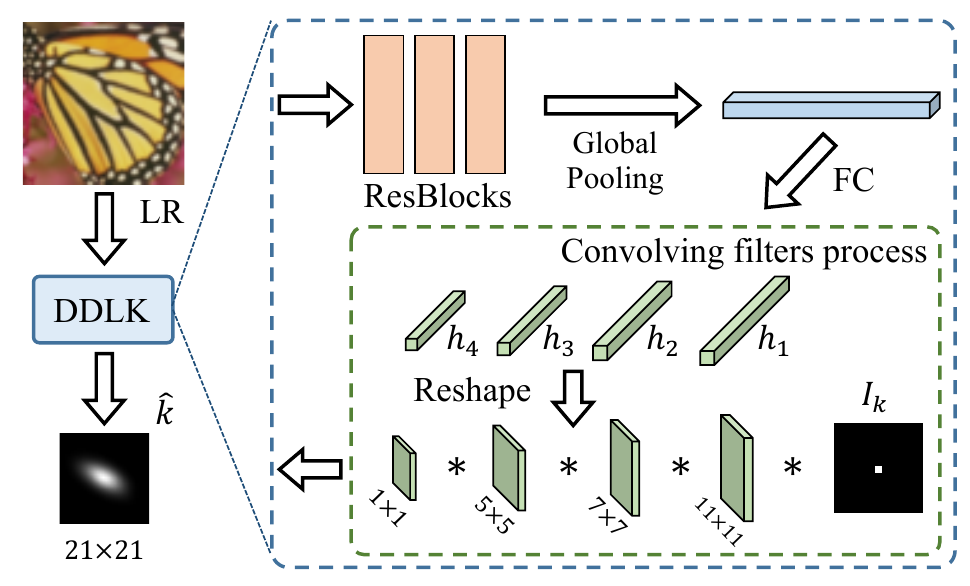}
\end{center}
\caption{Architecture of the dynamic deep linear kernel.}
\label{fig:DDLK}
\end{figure}

\subsection{Dynamic Deep Linear Kernel}


Following the reformation, we start our blind SR method from the kernel estimation. A straightforward solution is to adopt a regression network to estimate kernel ${\bf \hat k}$ by minimizing the L1 difference w.r.t the new ground-truth blur kernel ${\bf k}_l$ in Eq.~(\ref{eq:derive_stable_k}). We argue such a single layer kernel (all weights of estimated kernel equal to the ground-truth kernel) estimation is in general difficult and unstable due to the highly non-convex of the blind SR problem~\cite{bell2019blind}, leading to kernel mismatch and performance drop~\cite{gu2019blind,luo2020unfolding}. Instead, we propose an image-specific dynamic deep linear kernel (DDLK) which consists of a sequence of linear convolution layers without activations. Theoretically, deep linear networks have infinitely equivalent global minimas~\cite{saxe2013exact,kawaguchi2016deep,bell2019blind}, which allow us to find many different filter parameters to achieve the same correct solution. Moreover, since no non-linearity is used in the network, we can analytically collapse a deep linear kernel as a single layer kernel.

\begin{figure*}[t]
\setlength{\abovecaptionskip}{-0.in}
\setlength{\belowcaptionskip}{0.in}
\begin{center}
\includegraphics[width=1.\linewidth]{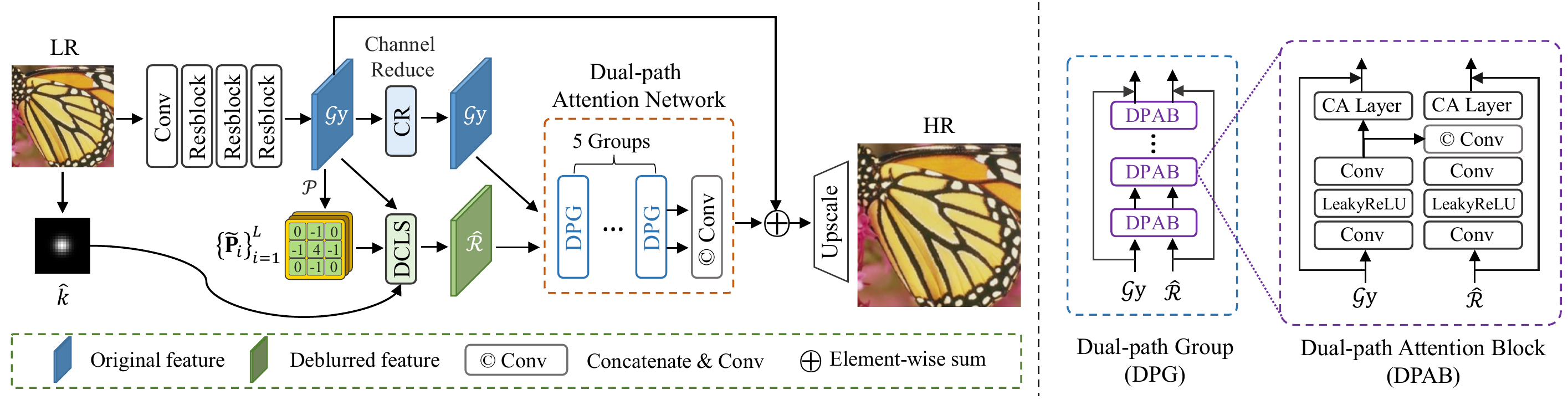}
\end{center}
\caption{The overview architecture of the proposed method. Given an LR image $\bf y$, we first estimate the degradation kernel $\hat{\bf k}$, and involve it in the deep constrained least squares (DCLS) convolution in the feature domain. The deblurred features $\widehat {\cal R}$ are then concatenated with primitive features ${\cal G}{\bf y}$ to restore the clean HR image $\bf x$ through a dual-path attention network (DPAN).}
\label{fig:pipline}
\end{figure*}

Fig.~\ref{fig:DDLK} depicts an example of estimating 4 layers dynamic deep linear kernel. The filters are set to $11\times11$, $7\times7$, $5\times5$ and $1\times1$, which make the receptive field to be $21\times21$. We first generate the filters of each layer based on the LR image, and explicitly sequentially convolve all filters into a single narrow kernel with stride 1. Mathematically, let ${\bf h}_i$ represent the $i$-th layer filter, we can get a single layer kernel following
\begin{equation}
\centering
    {\bf \hat k} = {\bf I_k} \ast {\bf h}_1 \ast {\bf h}_2 \ast \cdots \ast  {\bf h}_r
\label{eq:convolve_k}
\end{equation}
where $r$ is the number of linear layers, ${\bf I_k}$ is an identity kernel. As an empirically prior, we also constrain the kernel ${\bf \hat k}$ sum up to 1. The kernel estimation network can be optimized by minimizing the L1 loss between estimated kernel ${\bf \hat k}$ and new ground-truth blur kernel ${\bf k}_l$ from Eq.~(\ref{eq:derive_stable_k}).

\subsection{Deep Constrained Least Squares}

Our goal is to restore HR image based on LR image and estimated kernel ${\bf \hat k}$ according to the new degradation model (Eq. (\ref{eq:new_degrad})). 
Considering a group of feature extracting linear layers $\{{\cal G}_i\}_{i=1}^L$ provided to the LR image, we can rewrite Eq. (\ref{eq:new_degrad}) in the feature space, given by
\begin{equation}
\centering
    {\cal G}_i{\bf y} = {\bf \hat k} {\cal G}_i{\bf x}_{\downarrow_s} + {\cal G}_i{\bf n}.
\label{eq:fea_degrad}
\end{equation}
Let ${\widehat {\cal R}_i}$ be the sought after deblurred feature corresponding to ${\cal G}_i{\bf x}_{\downarrow_s}$. To solve Eq. (\ref{eq:fea_degrad}), we minimize the following criterion function
\begin{equation}
\centering
    {\cal C} = || \nabla \widehat{\mathcal R}_i ||^2, \; s.t. \ \ || {\cal G}_i{\bf y} - {\bf \hat k}{\widehat {\cal R}_i} ||^2 = || {\cal G}_i{\bf n} ||^2
\label{eq:cri_func}
\end{equation}
where the $\nabla$ is a smooth filter which can be denoted by $\bf P$. Then we introduce the Lagrange function, defined by
\begin{equation}
\centering
    \min_{{\widehat {\cal R}_i}} \left [ || {\bf P} \widehat{\mathcal R}_i ||^2 + \lambda \left ( || {\cal G}_i{\bf y} - {\bf \hat k}{\widehat {\cal R}_i} ||^2 - || {\cal G}_i{\bf n} ||^2 \right) \right],
\label{eq:lagrange_func}
\end{equation}
where $\lambda$ is the Lagrange multiplier. Computing the derivative of Eq.~(\ref{eq:lagrange_func}) with respect to ${\widehat {\cal R}_i}$ and setting it to zero:
\begin{equation}
\centering
    \left( \lambda {\bf \hat k}^{\mathsf{T}}{\bf \hat k} + {\bf P}^{\mathsf{T}}{\bf P} \right) \widehat{\mathcal R}_i - \lambda {\bf \hat k}^{\mathsf{T}} {\cal G}_i{\bf y} = 0.
\label{eq:lagrange_derive}
\end{equation}
We can obtain the clear features as
\begin{equation}
\centering
    \widehat{\mathcal R}_i = {\cal H} {\cal G}_i{\bf y}.
\label{eq:dcls}
\end{equation}
where ${\cal H}_i$ denotes the deep constrained least squares deconvolution (DCLS) operator, given by
\begin{equation}
\centering
    {\cal H} = {\cal F}^{-1} \left( \frac{\overline{{\cal F}({\bf \hat k})}}{\overline{{\cal F}({\bf \hat k})}{\cal F}({\bf \hat k}) + \frac{1}{\lambda }\overline{{\cal F}({\bf P})}{\cal F}({\bf P})} \right).
\label{eq:h_simple}
\end{equation}
Different from in the standard image space (e.g. RGB), smooth filter ${\bf P}$ and variable $\lambda$ in Eq.~(\ref{eq:h_simple}) might be inconsistent in the feature space. Alternatively, we predict a group of smooth filters with implicit Lagrange multiplier for different channels through a neural network $\cal P$:
\begin{equation}
\centering
    \{\tilde {\bf P}_i\}_{i=1}^L=\{{\cal P}({\cal G}_i{\bf y})\}_{i=1}^L.
\label{eq:pred_P}
\end{equation}
Then the feature-specific operator ${\cal H}_i$ can be define by 
\begin{equation}
\centering
    {\cal H}_i = {\cal F}^{-1} \left( \frac{\overline{{\cal F}({\bf \hat k})}}{\overline{{\cal F}({\bf \hat k})}{\cal F}({\bf \hat k}) + \overline{{\cal F}(\tilde{\bf P}_i)}{\cal F}(\tilde{\bf P}_i)} \right).
\label{eq:h_deep}
\end{equation}
Now we can obtain the clear features by Eq.~(\ref{eq:dcls}) and Eq.~(\ref{eq:h_deep}). 

It is worth to note that a deep neural network (DNN) can be locally linear~\cite{dong2020deep,lee2018towards,montufar2014number}, thus we could apply DNN as ${\cal G}_i$ to extract useful features in Eq.~(\ref{eq:fea_degrad}). In addition, the consequent artifacts or errors can be compensated by the following dual-path attention module.



\begin{table*}[ht]
\setlength{\abovecaptionskip}{0.05in}
\setlength{\belowcaptionskip}{-0.1in}
\centering
\resizebox{0.91\linewidth}{!}{
\begin{tabular}{cccccccccccc}
\toprule
\multirow{2}{*}{Method}& \multirow{2}{*}{Scale} & \multicolumn{2}{c}{Set5~\cite{bevilacqua2012low}}              & \multicolumn{2}{c}{Set14~\cite{zeyde2010single}}     & \multicolumn{2}{c}{BSD100~\cite{martin2001database}}      & \multicolumn{2}{c}{Urban100~\cite{huang2015single}}          & \multicolumn{2}{c}{Manga109~\cite{matsui2017sketch}}                     \\ 
& \multirow{-2}{*}{}  & PSNR  & SSIM         & PSNR  & SSIM      & PSNR  & SSIM         & PSNR  & SSIM      & PSNR  & SSIM   \\ \midrule
Bicubic & \multirow{9}{*}{\tabincell{c}{x2}}  & 28.82 &0.8577   &26.02 &0.7634  &25.92 &0.7310	&23.14 &0.7258  &25.60 &0.8498  \\
CARN~\cite{ahn2018fast} & & 30.99 &0.8779	&28.10  &0.7879	   & 26.78 & 0.7286	    &25.27 &0.7630	    & 26.86 &0.8606  \\
Bicubic+ZSSR~\cite{shocher2018zero} & & 31.08 &0.8786	&28.35  &0.7933	   & 27.92 & 0.7632	    &25.25 &0.7618	    & 28.05 &0.8769  \\
Deblurring~\cite{pan2017deblurring}+CARN~\cite{shocher2018zero} & & 24.20 &0.7496	&21.12  &0.6170	   & 22.69 & 0.6471	    &18.89 &0.5895	    & 21.54 &0.7946  \\
CARN~\cite{shocher2018zero}+Deblurring~\cite{pan2017deblurring} & & 31.27 &0.8974	&29.03  &0.8267	   & 28.72 & 0.8033	    &25.62 &0.7981	    & 29.58 &0.9134  \\
IKC~\cite{gu2019blind} & & 37.19 &0.9526	&32.94  &0.9024	   & 31.51 & 0.8790	    &29.85 &0.8928	    & 36.93 &0.9667  \\
DANv1~\cite{luo2020unfolding} & & 37.34 &0.9526 	&33.08  &0.9041	   & 31.76 & 0.8858	    &30.60 &0.9060	    & 37.23 &0.9710  \\

DANv2~\cite{luo2021endtoend} &  & {\color[HTML]{3531FF} 37.60} & {\color[HTML]{3531FF} 0.9544}     & {\color[HTML]{3531FF} 33.44} & {\color[HTML]{3531FF} 0.9094}    & {\color[HTML]{3531FF} 32.00} & {\color[HTML]{3531FF} 0.8904}   & {\color[HTML]{3531FF} 31.43} & {\color[HTML]{3531FF} 0.9174}    & {\color[HTML]{3531FF} 38.07} & {\color[HTML]{3531FF} 0.9734}  \\ 

DCLS(Ours) &  & {\color[HTML]{FE0000} 37.63} & {\color[HTML]{FE0000} 0.9554}     & {\color[HTML]{FE0000} 33.46} & {\color[HTML]{FE0000} 0.9103}    & {\color[HTML]{FE0000} 32.04} & {\color[HTML]{FE0000} 0.8907}   & {\color[HTML]{FE0000} 31.69} & {\color[HTML]{FE0000} 0.9202}    & {\color[HTML]{FE0000} 38.31} & {\color[HTML]{FE0000} 0.9740}  \\

\midrule

Bicubic & \multirow{9}{*}{\tabincell{c}{x3}}  & 26.21 &0.7766   &24.01 &0.6662  &24.25 &0.6356	&21.39 &0.6203  &22.98 &0.7576  \\
CARN~\cite{ahn2018fast} & & 27.26 &0.7855	&25.06  &0.6676	   & 25.85 & 0.6566	    &22.67 &0.6323	    & 23.85 &0.7620  \\
Bicubic+ZSSR~\cite{shocher2018zero} & & 28.25 &0.7989	&26.15  &0.6942	   & 26.06 & 0.6633	    &23.26 &0.6534	    & 25.19 &0.7914  \\
Deblurring~\cite{pan2017deblurring}+CARN~\cite{shocher2018zero} & & 19.05 &0.5226	&17.61  &0.4558	   & 20.51 & 0.5331	    &16.72 &0.5895	    & 18.38 &0.6118  \\
CARN~\cite{shocher2018zero}+Deblurring~\cite{pan2017deblurring} & & 30.31 &0.8562	&27.57  &0.7531	   & 27.14 & 0.7152	    &24.45 &0.7241	    & 27.67 &0.8592  \\
IKC~\cite{gu2019blind} & & 33.06 &0.9146	&29.38  &0.8233	   & 28.53 & 0.7899	    &24.43 &0.8302	    & 32.43 &0.9316  \\
DANv1~\cite{luo2020unfolding} & & 34.04 &0.9199 	&30.09  &0.8287	   & 28.94 & 0.7919	    &27.65 &0.8352	    & 33.16 &0.9382  \\

DANv2~\cite{luo2021endtoend} &  & {\color[HTML]{3531FF} 34.12} & {\color[HTML]{3531FF} 0.9209}     & {\color[HTML]{3531FF} 30.20} & {\color[HTML]{3531FF} 0.8309}    & {\color[HTML]{3531FF} 29.03} & {\color[HTML]{3531FF} 0.7948}   & {\color[HTML]{3531FF} 27.83} & {\color[HTML]{3531FF} 0.8395}    & {\color[HTML]{3531FF} 33.28} & {\color[HTML]{3531FF} 0.9400}  \\ 

DCLS(Ours) &  & {\color[HTML]{FE0000} 34.21} & {\color[HTML]{FE0000} 0.9218}     & {\color[HTML]{FE0000} 30.29} & {\color[HTML]{FE0000} 0.8329}    & {\color[HTML]{FE0000} 29.07} & {\color[HTML]{FE0000} 0.7956}   & {\color[HTML]{FE0000} 28.03} & {\color[HTML]{FE0000} 0.8444}    & {\color[HTML]{FE0000} 33.54} & {\color[HTML]{FE0000} 0.9414}  \\

\midrule

Bicubic & \multirow{9}{*}{\tabincell{c}{x4}}  & 24.57 &0.7108   &22.79 &0.6032  &23.29 &0.5786	&20.35 &0.5532  &21.50 &0.6933  \\
CARN~\cite{ahn2018fast} & & 26.57 &0.7420	&24.62  &0.6226	   & 24.79 & 0.5963	    &22.17 &0.5865	    & 21.85 &0.6834  \\
Bicubic+ZSSR~\cite{shocher2018zero} & & 26.45 &0.7279	&24.78  &0.6268	   & 24.97 & 0.5989	    &22.11 &0.5805	    & 23.53 &0.7240  \\
Deblurring~\cite{pan2017deblurring}+CARN~\cite{shocher2018zero} & & 18.10 &0.4843	&16.59  &0.3994	   & 18.46 & 0.4481	    &15.47 &0.3872	    & 16.78 &0.5371  \\
CARN~\cite{shocher2018zero}+Deblurring~\cite{pan2017deblurring} & & 28.69 &0.8092	&26.40  &0.6926	   & 26.10 & 0.6528	    &23.46 &0.6597	    & 25.84 &0.8035  \\
IKC~\cite{gu2019blind} & & 31.67 &0.8829	&28.31  &0.7643	   & 27.37 & 0.7192	    &25.33 &0.7504	    & 28.91 &0.8782  \\
DANv1~\cite{luo2020unfolding} & & 31.89 &0.8864 	&28.42  &0.7687	   & 27.51 & 0.7248	    &25.86 &0.7721	    &{\color[HTML]{3531FF} 30.50} & {\color[HTML]{3531FF} 0.9037}  \\

DANv2~\cite{luo2021endtoend} &  & {\color[HTML]{3531FF} 32.00} & {\color[HTML]{3531FF} 0.8885}     & {\color[HTML]{3531FF} 28.50} & {\color[HTML]{3531FF} 0.7715}    & {\color[HTML]{3531FF} 27.56} & {\color[HTML]{3531FF} 0.7277}   & {\color[HTML]{3531FF} 25.94} & {\color[HTML]{3531FF} 0.7748}    & 30.45 & {\color[HTML]{3531FF} 0.9037}  \\ 

AdaTarget~\cite{jo2021adatarget}& & 31.58 &0.8814 	&28.14  &0.7626	   & 27.43 & 0.7216	    &25.72 &0.7683	    & 29.97 &0.8955  \\

DCLS(Ours) &  & {\color[HTML]{FE0000} 32.12} & {\color[HTML]{FE0000} 0.8890}     & {\color[HTML]{FE0000} 28.54} & {\color[HTML]{FE0000} 0.7728}    & {\color[HTML]{FE0000} 27.60} & {\color[HTML]{FE0000} 0.7285}   & {\color[HTML]{FE0000} 26.15} & {\color[HTML]{FE0000} 0.7809}    & {\color[HTML]{FE0000} 30.86} & {\color[HTML]{FE0000} 0.9086}  \\

\bottomrule
\end{tabular}
}
\caption{Quantitative comparison on datasets with \textit{Gaussian8} kernels. The best two results are marked in {\color{red}{red}} and {\color{blue}{blue}} colors, respectively.}
\label{table:iso_cmp}
\end{table*}

\begin{table*}[ht]
\setlength{\abovecaptionskip}{0.05in}
\setlength{\belowcaptionskip}{-0.05in}
\centering
\resizebox{0.91\linewidth}{!}{
\begin{tabular}{cccccccccccc}
\toprule
\multirow{2}{*}{Method $\times$4}& \multirow{2}{*}{Noise level} & \multicolumn{2}{c}{Set5~\cite{bevilacqua2012low}}              & \multicolumn{2}{c}{Set14~\cite{zeyde2010single}}     & \multicolumn{2}{c}{BSD100~\cite{martin2001database}}      & \multicolumn{2}{c}{Urban100~\cite{huang2015single}}          & \multicolumn{2}{c}{Manga109~\cite{matsui2017sketch}}                     \\ 
& \multirow{-2}{*}{}  & PSNR  & SSIM         & PSNR  & SSIM      & PSNR  & SSIM         & PSNR  & SSIM      & PSNR  & SSIM   \\ \midrule

Bicubic+ZSSR~\cite{shocher2018zero} &\multirow{5}{*}{\tabincell{c}{15}} & 23.32 &0.4868	&22.49  &0.4256	   & 22.61 & 0.3949	    &20.68 &0.3966	    & 22.04 &0.4952  \\
IKC~\cite{gu2019blind} & & 26.89 &0.7671	&25.28  &0.6483	   & 24.93 & 0.6019	    &22.94 &0.6362	    & 25.09 &0.7819  \\
DANv1~\cite{luo2020unfolding} & & 26.95 &0.7711 	&25.27  &0.6490	   & {\color[HTML]{3531FF} 24.95} & {\color[HTML]{3531FF} 0.6033}	    &23.00 &0.6407	    &25.29 & 0.7879  \\

DANv2~\cite{luo2021endtoend} &  & {\color[HTML]{3531FF} 26.97} & {\color[HTML]{3531FF} 0.7726}     & {\color[HTML]{3531FF} 25.29} & {\color[HTML]{3531FF} 0.6497}    & {\color[HTML]{3531FF} 24.95} & 0.6025   & {\color[HTML]{3531FF} 23.03} & {\color[HTML]{3531FF} 0.6429}    & {\color[HTML]{3531FF} 25.32} & {\color[HTML]{3531FF} 0.7896}  \\ 

DCLS(Ours) &  & {\color[HTML]{FE0000} 27.14} & {\color[HTML]{FE0000} 0.7775}     & {\color[HTML]{FE0000} 25.37} & {\color[HTML]{FE0000} 0.6516}    & {\color[HTML]{FE0000} 24.99} & {\color[HTML]{FE0000} 0.6043}   & {\color[HTML]{FE0000} 27.13} & {\color[HTML]{FE0000} 0.6500}    & {\color[HTML]{FE0000} 25.57} & {\color[HTML]{FE0000} 0.7969}  \\

\midrule

Bicubic+ZSSR~\cite{shocher2018zero} &\multirow{5}{*}{\tabincell{c}{30}} & 19.77 &0.2938	&19.36  &0.2534	   & 19.43 & 0.2308	    &18.32 &0.2450	    & 19.25 &0.3046  \\
IKC~\cite{gu2019blind} & & 25.27 &0.7154	&24.15  &0.6100	   & {\color[HTML]{3531FF} 24.06} & 0.5674	    &22.11 &0.5969	    & 23.80 &0.7438  \\
DANv1~\cite{luo2020unfolding} & & 25.32 & {\color[HTML]{3531FF} 0.7276} 	&24.15  & {\color[HTML]{3531FF}0.6138}	   & 24.04 & 0.5678	    &22.08 &0.5977	    &23.82 & 0.7442  \\

DANv2~\cite{luo2021endtoend} &  & {\color[HTML]{3531FF} 25.36} &  0.7264     & {\color[HTML]{3531FF} 24.16} & 0.6121    & {\color[HTML]{3531FF} 24.06} &  {\color[HTML]{3531FF} 0.5690}   & {\color[HTML]{3531FF} 22.14} & {\color[HTML]{3531FF} 0.6014}    & {\color[HTML]{3531FF} 23.87} & {\color[HTML]{3531FF} 0.7489}  \\ 

DCLS(Ours) &  & {\color[HTML]{FE0000} 25.49} & {\color[HTML]{FE0000} 0.7323}     & {\color[HTML]{FE0000} 24.23} & {\color[HTML]{FE0000} 0.6131}    & {\color[HTML]{FE0000} 24.09} & {\color[HTML]{FE0000} 0.5696}   & {\color[HTML]{FE0000} 22.37} & {\color[HTML]{FE0000} 0.6119}    & {\color[HTML]{FE0000} 24.21} & {\color[HTML]{FE0000} 0.7582}  \\

\bottomrule
\end{tabular}
}
\caption{Quantitative comparison on various noisy datasets. The best one marks in {\color{red}{red}} and the second best are in {\color{blue}{blue}}.}
\label{table:iso_noise_cmp}
\end{table*}

\subsection{Dual-Path Attention Network}

Unlike previous works~\cite{gu2019blind,luo2021endtoend} in which the dual-path structures are only used to concatenate the stretched kernel with blurred features, we propose to utilize primitive blur features as additive path to compensate the artifacts and errors introduced by the estimated kernel, known as dual-path attention network (DPAN).
DPAN is composed of several groups of dual-path attention blocks (DPAB), it receives both deblurred features $\widehat{\mathcal R}$ and primitive features ${\cal G}{\bf y}$. 
The right of Fig.~\ref{fig:pipline} illustrates the architecture of DPAB. 

Since the additive path of processing ${\cal G}{\bf y}$ is independently updated and used to concatenate with $\widehat{\mathcal R}$ to provide primary information to refine the deconvolved features. We can reduce its channels to accelerate training and inference, as the channel reduction (CR) operation illustrated in left of Fig.~\ref{fig:pipline}. Moreover, on the deconvolved feature path, we apply the channel attention layer~\cite{Zhang_2018_ECCV} after aggregating original features. In addition, we add a residual connection for each path on all groups and blocks. 
The pixelshuffle~\cite{huang2009multi} is used as the upscale module. We can jointly optimize the SR network and kernel estimation network as follows:
\begin{equation}
\centering
    {\cal L} =  l_1({\bf \hat k}, {\bf k}_l; \theta_k) + l_1({\bf \hat x}, {\bf x}; \theta_g)
\label{eq:loss}
\end{equation}
where $\theta_k$ and $\theta_g$ are the parameters of kernel estimation network and DCLS reconstruction network, respectively.

\section{Experiments}

\subsection{Datasets and Implementation Details}
Following previous works~\cite{gu2019blind,luo2020unfolding}, 3450 2K HR images from DIV2K~\cite{agustsson2017ntire} and Flickr2K~\cite{timofte2017ntire} are collected as the training dataset. And we synthesize corresponding LR images with specific degradation kernel settings (e.g., isotropic/anisotropic Gaussian) using Eq. ~(\ref{eq:class_degrad}). The proposed method is evaluated by PSNR and SSIM~\cite{wang2004image} on only the luminance channel of the SR results (YCbCr space).

\noindent \textbf{Isotropic Gaussian kernels.} 
Firstly, we conduct blind SR experiments on isotropic Gaussian kernels following the setting in~\cite{gu2019blind}. Specifically, the kernel sizes are fixed to 21 $\times$ 21. In training, we uniformly sample the kernel width from range [0.2, 2.0], [0.2, 3.0] and [0.2, 4.0] for SR scale factors 2, 3 and 4, respectively. For testing, we use \textit{Gaussian8}~\cite{gu2019blind} kernel setting to generate evaluation dataset from five widely used benchmarks: Set5~\cite{bevilacqua2012low}, Set14~\cite{zeyde2010single}, BSD100~\cite{martin2001database}, Urban100~\cite{huang2015single} and Manga109~\cite{matsui2017sketch}. \textit{Gaussian8} uniformly chooses 8 kernels from range [0.80, 1.60], [1.35, 2.40] and [1.80, 3.20] for scale factors 2, 3 and 4, repectively. The LR images are obtained by blurring and downsampling the HR images with selected kernels.

\noindent \textbf{Anisotropic Gaussian kernels.}
We also conduct experiments on anisotropic Gaussian kernels following the setting in~\cite{bell2019blind}. The kernel size is set to 11 $\times$ 11 and 31 $\times$ 31 for scale factors 2 and 4, respectively. During training, the anisotropic Gaussian kernels for degradation are generated by randomly selecting kernel width from range (0.6, 5) and rotating from range [-$\pi$, $\pi$]. We also apply uniform multiplicative noise and normalize it to sum to one. For evaluation, we use the DIV2KRK dataset proposed in~\cite{bell2019blind}.

\begin{figure}
\setlength{\abovecaptionskip}{0.05in}
\setlength{\belowcaptionskip}{-0.05in}
\centering
	\begin{minipage}[t]{0.49\linewidth}
		\centering
		\includegraphics[width=1.64in]{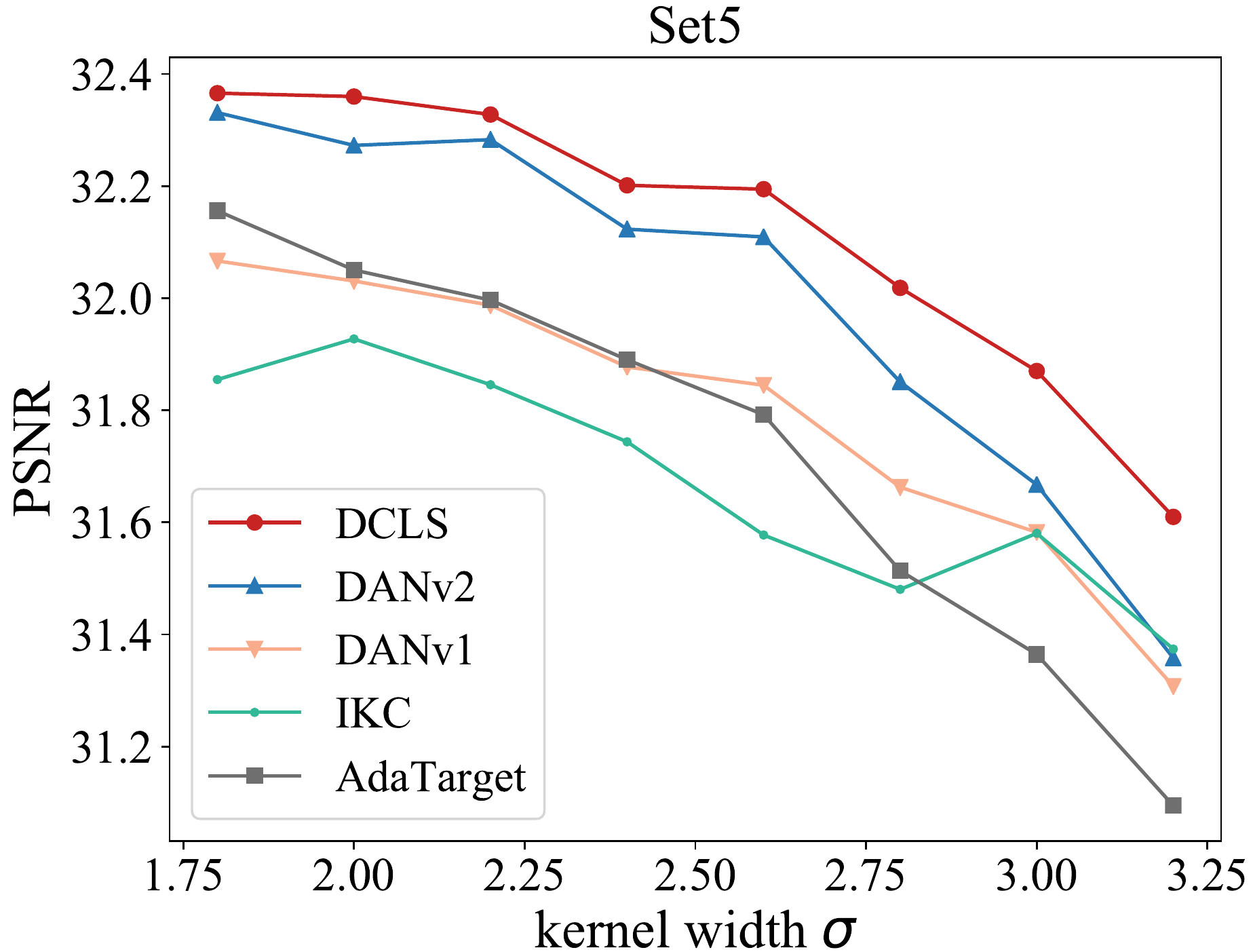}
	\end{minipage}
	\begin{minipage}[t]{0.49\linewidth}
		\centering
		\includegraphics[width=1.57in]{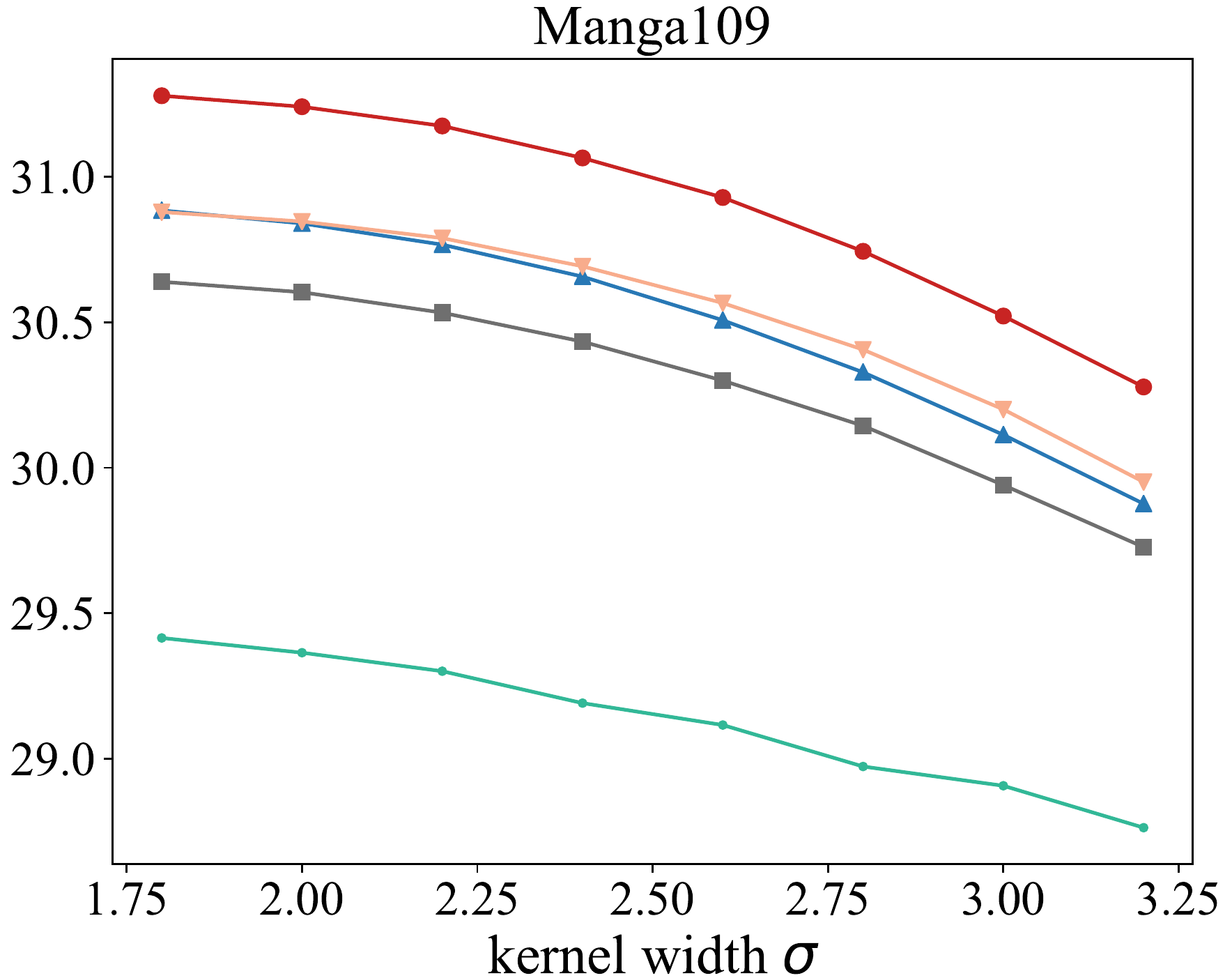}
	\end{minipage}
	\caption{The PSNR performance curves on Set5 and Manga109 of scale factor 4. The kernel width $\sigma$ are set from 1.8 to 3.2. }
	\label{fig:cmp_psnr}
\end{figure}

\begin{figure}[t]
\setlength{\abovecaptionskip}{-0.1in}
\setlength{\belowcaptionskip}{-0.1in}
\begin{center}
\includegraphics[width=.98\linewidth]{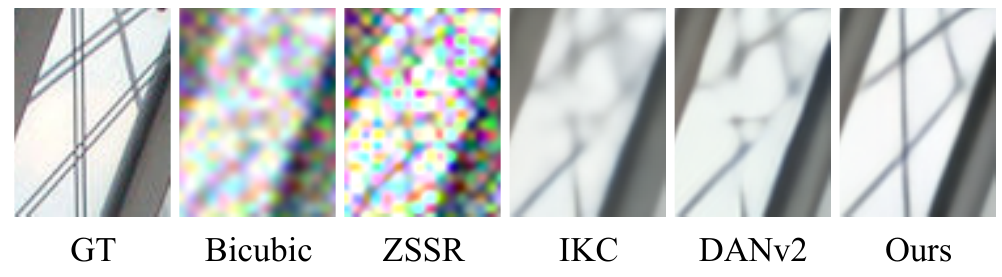}
\end{center}
\caption{Visual results of \textit{Img 33} from Urban100.}
\label{fig:cmp_noise}
\end{figure}

\noindent \textbf{Implementation details.}
For all experiments, we use 5 dual-path groups, each containing 10 DPABs with 64 channels. The batch sizes are set to 64 and the LR patch sizes are 64 $\times$ 64. We use Adam~\cite{kingma2014adam} optimizer with $\beta_1=0.9$ and $\beta_2=0.99$. All models are trained on 4 RTX2080Ti GPUs with $5 \times 10^5$ iterations. The initial learning rate is set to $4 \times 10^{-4}$ and decayed by half at every $2 \times 10^{-4}$ iterations. We also augment the training data with random horizontal flips and 90 degree rotations.

\begin{table}[t]
\setlength{\abovecaptionskip}{0.05in}
\setlength{\belowcaptionskip}{-0.05in}
\centering
\resizebox{1.0\linewidth}{!}{
\begin{tabular}{lcccc}
\toprule
\multirow{3}{*}{Method}     & \multicolumn{4}{c}{DIV2KRK~\cite{bell2019blind}}     \\ 

& \multicolumn{2}{c}{$\times$2}              & \multicolumn{2}{c}{$\times$4}    \\ 
& PSNR  & SSIM         & PSNR  & SSIM  \\ \midrule
Bicubic & 28.73 &0.8040   &25.33 &0.6795  \\
Bicubic+ZSSR~\cite{shocher2018zero} & 29.10 &0.8215   &25.61 &0.6911  \\
EDSR~\cite{lim2017enhanced} & 29.17 &0.8216   &25.64 &0.6928  \\
RCAN~\cite{zhang2018image} & 29.20 &0.8223   &25.66 &0.6936  \\
DBPN~\cite{haris2018deep} & 29.13 & 0.8190   &25.58 &0.6910  \\
DBPN~\cite{haris2018deep}+Correction~\cite{hussein2020correction} & 30.38 &0.8717   &26.79 &0.7426  \\

KernelGAN~\cite{bell2019blind}+SRMD~\cite{zhang2018learning}  & 29.57 &0.8564   &27.51 &0.7265  \\
KernelGAN~\cite{bell2019blind}+ZSSR~\cite{shocher2018zero}  & 30.36 & 0.8669   &26.81 &0.7316  \\

IKC~\cite{gu2019blind} & -  & -   &27.70 &0.7668  \\
DANv1~\cite{luo2020unfolding} & 32.56 & 0.8997      &27.55 &0.7582  \\
DANv2~\cite{luo2021endtoend} & {\color[HTML]{3531FF} 32.58} & {\color[HTML]{3531FF} 0.9048}     & {\color[HTML]{3531FF} 28.74} & {\color[HTML]{3531FF} 0.7893}    \\ 
AdaTarget~\cite{jo2021adatarget} & -  & -   &28.42 &0.7854  \\
KOALAnet~\cite{kim2021koalanet} & 31.89 & 0.8852     & 27.77 & 0.7637    \\

DCLS(Ours) & {\color[HTML]{FE0000} 32.75} & {\color[HTML]{FE0000} 0.9094}     & {\color[HTML]{FE0000} 28.99} & {\color[HTML]{FE0000} 0.7946}  \\
\bottomrule
\end{tabular}
}
\caption{Quantitative comparison on DIV2KRK. The best one marks in {\color{red}{red}} and the second best are in {\color{blue}{blue}}.}
\label{table:aniso_cmp}
\end{table}

\subsection{Comparison with State-of-the-arts}

\noindent \textbf{Evaluation with isotropic Gaussian kernels.} 
Following~\cite{gu2019blind}, we evaluate our method on datasets synthesized by \textit{Gaussian8} kernels. We compare our method with state-of-the-art blind SR approaches: ZSSR~\cite{shocher2018zero} (with bicubic kernel), IKC~\cite{gu2019blind}, DANv1~\cite{luo2020unfolding}, DANv2~\cite{luo2021endtoend} and AdaTarget~\cite{jo2021adatarget}. Following~\cite{gu2019blind}, we also conduct comparison with CARN~\cite{ahn2018fast} and its variants of performing blind deblurring method~\cite{pan2017deblurring} before and after CARN. For most methods, we use their official implementations and pre-trained models.

\begin{figure}[t]
\setlength{\abovecaptionskip}{-0.1in}
\setlength{\belowcaptionskip}{-0.05in}
\begin{center}
\includegraphics[width=.94\linewidth]{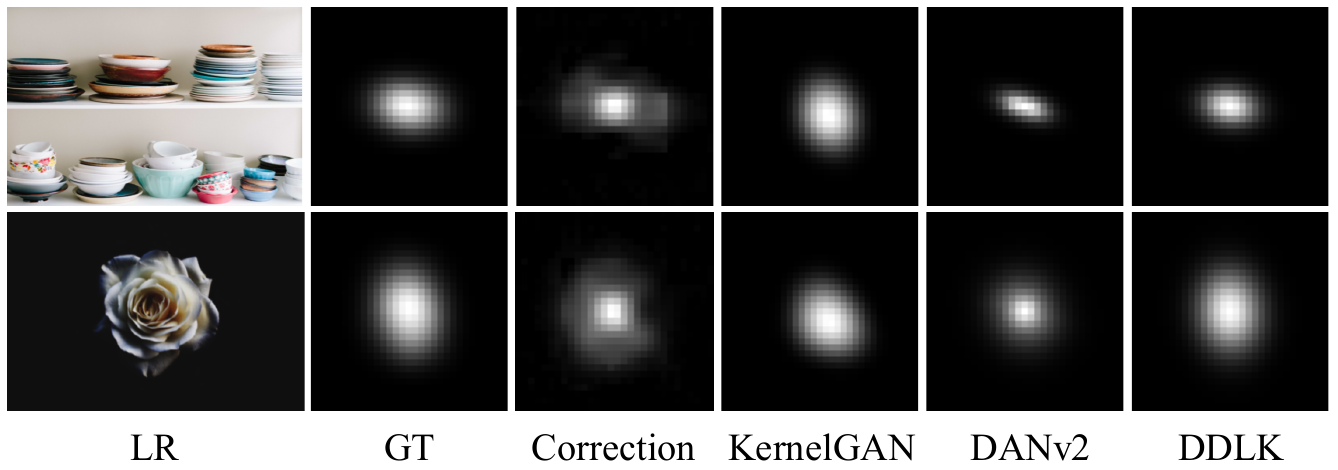}
\end{center}
\caption{Visual results of estimated kernels of \textit{Img 33} and \textit{Img 43} from DIV2KRK~\cite{bell2019blind} by various kernel estimation methods.}
\label{fig:cmp_k}
\end{figure}

\begin{table}[t]
\setlength{\abovecaptionskip}{0.05in}
\setlength{\belowcaptionskip}{-0.1in}
\centering
\resizebox{.95\linewidth}{!}{
\begin{tabular}{ccccc}
\toprule
DIV2KRK $\times4$   & KernelGAN   & CorrFilter      & DANv2  & DDLK       \\ \midrule

LR-PSNR $\uparrow$      &41.28      & 41.35      & 45.06      & \textbf{45.27}   \\
Kernel-MSE $\downarrow$     & 0.1518    & 0.1392       & 0.0817   & \textbf{0.0574}   \\

\bottomrule
\end{tabular}
}
\caption{Quantitative evaluation on the performance of DDLK.}
\label{table:cmp_k}
\end{table}

\begin{figure*}[t]
\setlength{\abovecaptionskip}{-0.1in}
\setlength{\belowcaptionskip}{-0.1in}
\begin{center}
\includegraphics[width=.999\linewidth]{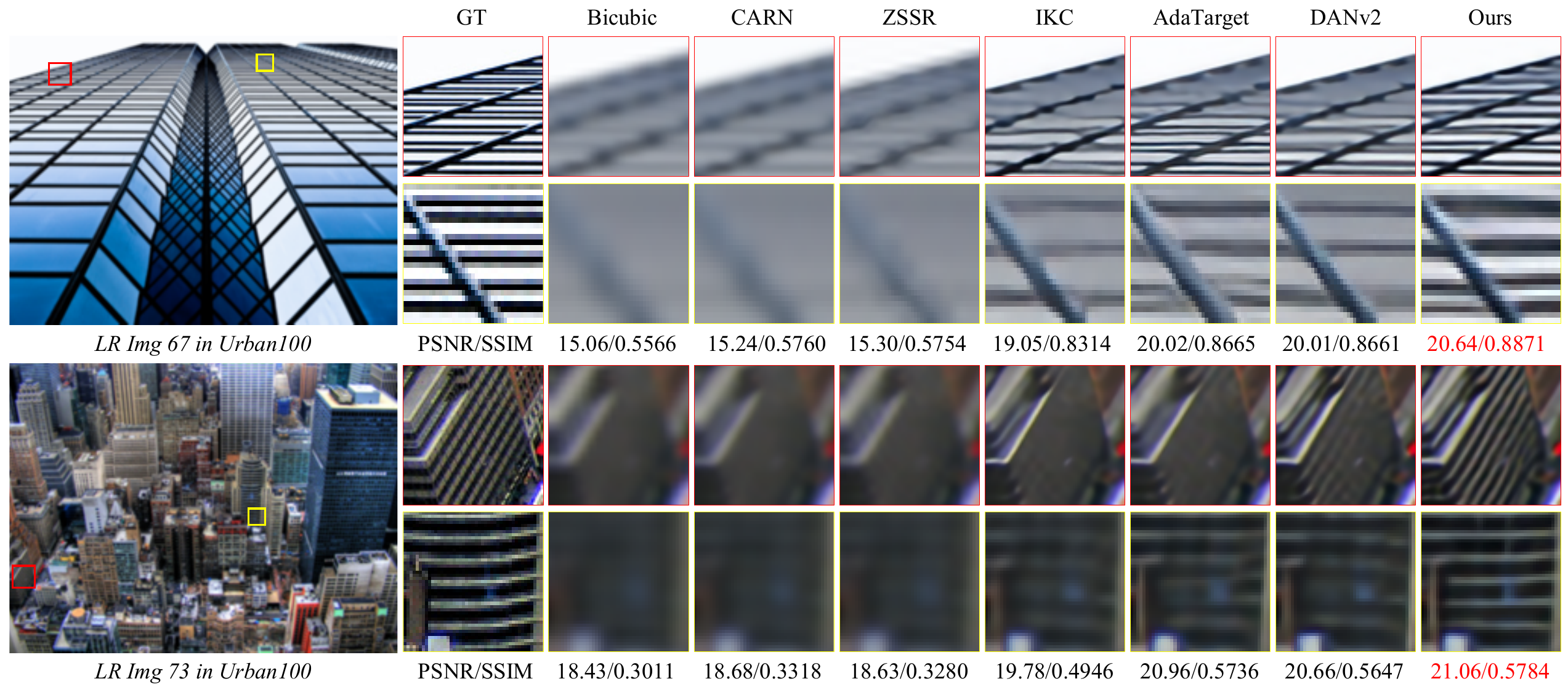}
\end{center}
\caption{Visual results of \textit{Img 67} and \textit{Img 73} in Urban100~\cite{huang2015single}, for scale factor 4 and kernel width 2.6. Best viewed in color.}
\label{fig:cmp_iso}
\end{figure*}

\begin{figure*}[t]
\setlength{\abovecaptionskip}{-0.1in}
\setlength{\belowcaptionskip}{-0.1in}
\begin{center}
\includegraphics[width=.999\linewidth]{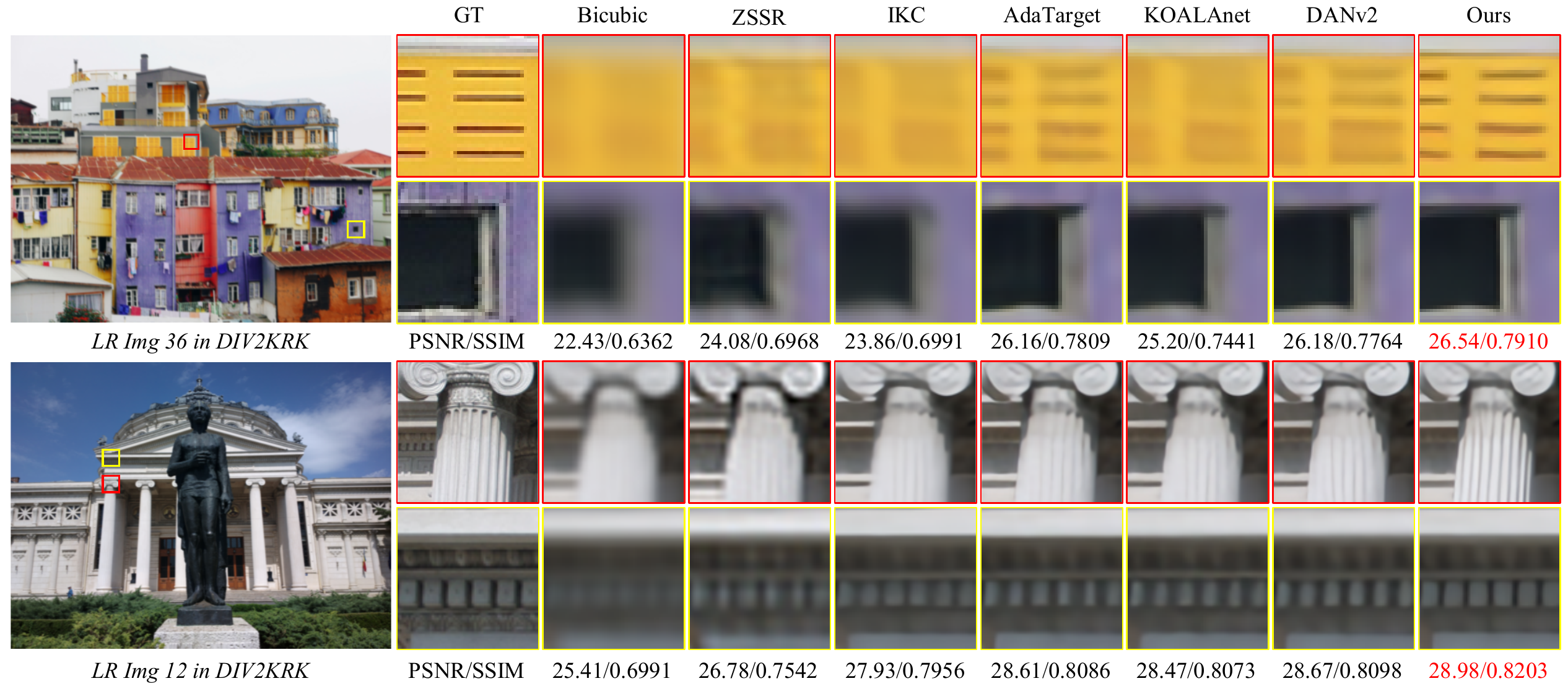}
\end{center}
\caption{Visual results of \textit{Img 36} and \textit{Img 12} in DIV2KRK~\cite{bell2019blind}, for scale factor of 4. Best viewed in color.}
\label{fig:cmp_aniso}
\end{figure*}

The quantitative results are shown in \Cref{table:iso_cmp}. It is obvious that our method leads to the best performance over all datasets. The bicubic SR model CARN suffers severe performance drop with \textit{Gaussian8} which deviates from the predefined bicubic kernel. Performing deblurring on the super-resolved image can improve the results. ZSSR achieves better performance compared with non-blind SR method but is limited by the image-specific network design (cannot utilize abundant training data). AdaTarget can improve image quality but is still inferior to that of blind SR methods. IKC and DAN are two-step blind SR methods and can largely improve the results. However, both of them predict kernel embedding and directly involve it into the network, which damages the spatial relation of the kernel and thus performs inferior to our method. We also provide the comparison of PSNR values on different datasets with blur kernels width from 1.8 to 3.2 as shown in Fig.~\ref{fig:cmp_psnr}. DCLS performs the best result over all different kernel widths. The qualitative results shown in Fig.~\ref{fig:cmp_iso} illustrate that DCLS can produce clear and pleasant SR images. Furthermore, we conduct an experiment of super-resolving images with additional noise. As shown in~\Cref{table:iso_noise_cmp} and Fig.~\ref{fig:cmp_noise}, DCLS still outperforms other methods over all datasets with different noise levels.

\noindent \textbf{Evaluation with anisotropic Gaussian kernels.} 
Degradation with anisotropic Gaussian kernels are more general and challenging. Similar to isotropic kernel, we firstly compare our method with SOTA blind SR approaches such as ZSSR~\cite{shocher2018zero}, IKC~\cite{gu2019blind}, DANv1~\cite{luo2020unfolding}, DANv2~\cite{luo2021endtoend}, AdaTarget~\cite{jo2021adatarget} and KOALAnet~\cite{kim2021koalanet}. We also compare DCLS with some SOTA bicubicly designed methods such as EDSR~\cite{lim2017enhanced}, RCAN~\cite{zhang2018image}, and DBPN~\cite{haris2018deep}. And we provide Correction~\cite{hussein2020correction} for DBPN. In addition, we combine a kernel estimation method (e.g. KernelGAN~\cite{bell2019blind}) with other non-blind SR methods, such as ZSSR~\cite{shocher2018zero} and SRMD~\cite{zhang2018learning}, as two-step solutions to solve blind SR. 

Table~\ref{table:aniso_cmp} shows the quantitative results on DIV2KRK~\cite{bell2019blind}. It can be seen that the proposed DCLS significantly improves the performance compared with other blind SR approaches. Note that ZSSR performs better when combined with KernelGAN, which indicates that good kernel estimation can help a lot. Recent SOTA blind SR methods such as IKC, DAN and KOALAnet can achieve remarkable accuracy in PSNR and SSIM. By applying an adaptive target to finetune the network, AdaTarget can perform comparably with SOTA blind methods. However, all of those methods are still inferior to the proposed DCLS. The visual results on DIV2KRK are shown in \cref{fig:cmp_aniso}. As we can see, the SR images produced by our method are much sharper and cleaner. We also provide the results of kernel estimation and downsampling HR image with estimated kernel in Fig.~\ref{fig:cmp_k} and Table~\ref{table:cmp_k}. Compared with previous image-specific methods such as KernelGAN~\cite{bell2019blind} and Correction Filter~\cite{hussein2020correction}, the dynamic deep linear kernel (DDLK) is more flexible and capable of producing accurate kernels.

\subsection{Analysis and Discussions}

\begin{table}[t]
\setlength{\abovecaptionskip}{0.05in}
\setlength{\belowcaptionskip}{-0.1in}
\centering
\resizebox{1.\linewidth}{!}{
\begin{tabular}{ccccc|cc}
\toprule
 \multirow{2}{*}{SLK}     & \multirow{2}{*}{DDLK}  & \multirow{2}{*}{\tabincell{c}{Stretching \\ Strategy}}   & \multirow{2}{*}{\tabincell{c}{DCLS \\ Deconv}}   & \multirow{2}{*}{DPAN}   & \multicolumn{2}{c}{DIV2KRK}     \\

    &      &     &      &   & PSNR  & SSIM    \\  \hline

 $\checkmark$     & -  & $\checkmark$     & -  & $\checkmark$  & 28.84 & 0.7921 \\
 -     & $\checkmark$       & $\checkmark$     &-  & $\checkmark$  & 28.86 & 0.7924    \\
$\checkmark$    & -       & -     & $\checkmark$  & $\checkmark$  & 28.94 & 0.7946    \\
 -     & $\checkmark$       & -    & $\checkmark$   & -  & 28.94 & 0.7938    \\
 -     & $\checkmark$       & -    & $\checkmark$   & $\checkmark$  & 28.99 & 0.7964    \\

\bottomrule
\end{tabular}
}
\caption{Ablation study on our vital components.}
\label{table:ablation}
\end{table}



\begin{table}[t]
\setlength{\abovecaptionskip}{0.05in}
\setlength{\belowcaptionskip}{-0.08in}
\centering
\resizebox{1.\linewidth}{!}{
\begin{tabular}{lcccccc}
\toprule
\multirow{2}{*}{Method}    & \multicolumn{2}{c}{$\rm Wiener_{Fea}$~\cite{dong2020deep}}       & \multicolumn{2}{c}{$\rm CLS_{Fea}$}       & \multicolumn{2}{c}{$\rm DCLS_{Fea}$}     \\
  & PSNR & SSIM     & PSNR & SSIM   & PSNR & SSIM   \\ \midrule
Set5      & 32.05 & 0.8878     & 31.98 & 0.8862     & \textbf{32.12} & \textbf{0.8890}    \\  
Set14     & 28.38 & 0.7709    & 28.29 & 0.7658     & \textbf{28.54} & \textbf{0.7728}    \\
BSD100    & 27.47 & 0.7238   & 27.48 & 0.7216     & \textbf{27.60} & \textbf{0.7285}    \\
Urban100    & 26.07 & 0.7775   & 26.03 & 0.7768     &\textbf{26.15} & \textbf{0.7809}    \\
Manga109    & 30.77 & 0.9069   & 30.65 & 0.9040     &\textbf{30.86} & \textbf{0.9086}    \\
DIV2KRK    & 28.77 & 0.7886   & 28.92 & 0.7921     &\textbf{28.99} & \textbf{0.7947}    \\

\bottomrule
\end{tabular}
}
\caption{Quantitative comparison on various datasets. $\rm Fea$ means applying deconvolution on the feature space.}
\label{table:fea-deconv}
\end{table}

\begin{table}[t]
\setlength{\abovecaptionskip}{0.05in}
\setlength{\belowcaptionskip}{-0.1in}
\centering
\resizebox{.99\linewidth}{!}{
\begin{tabular}{ccccc}
\toprule
DIV2KRK $\times$4    & $\rm Wiener_{RGB}$   & $\rm CLS_{RGB}$      & $\rm DCLS_{RGB}$  & $\rm DCLS_{Fea}$       \\ \midrule

PSNR      & 28.91     &28.90      & 28.94       & 28.99     \\
SSIM      & 0.7941    &0.7935     & 0.7941      & 0.7964      \\

\bottomrule
\end{tabular}
}
\caption{Quantitative results. $\rm RGB$ and $\rm Fea$ mean applying deconvolution in the RGB space and feature space, respectively.}
\label{table:rgb-deconv}
\end{table}

\noindent \textbf{Ablation Study.} 
We conduct ablation studies on vital components of our method: DPAN, DDLK and DCLS deconvolution. 
The quantitative results on DIV2KRK are exported in \Cref{table:ablation}. Note that the baseline model with DPAN eliminates artifacts from kernel and thus improves the result. And the DCLS deconvolution can further make use of the estimated kernel and high-level information from deep features to achieve a higher performance (+0.15dB from baseline). 

\begin{figure}[t]
\setlength{\abovecaptionskip}{-0.1in}
\setlength{\belowcaptionskip}{-0.05in}
\begin{center}
\includegraphics[width=1.\linewidth]{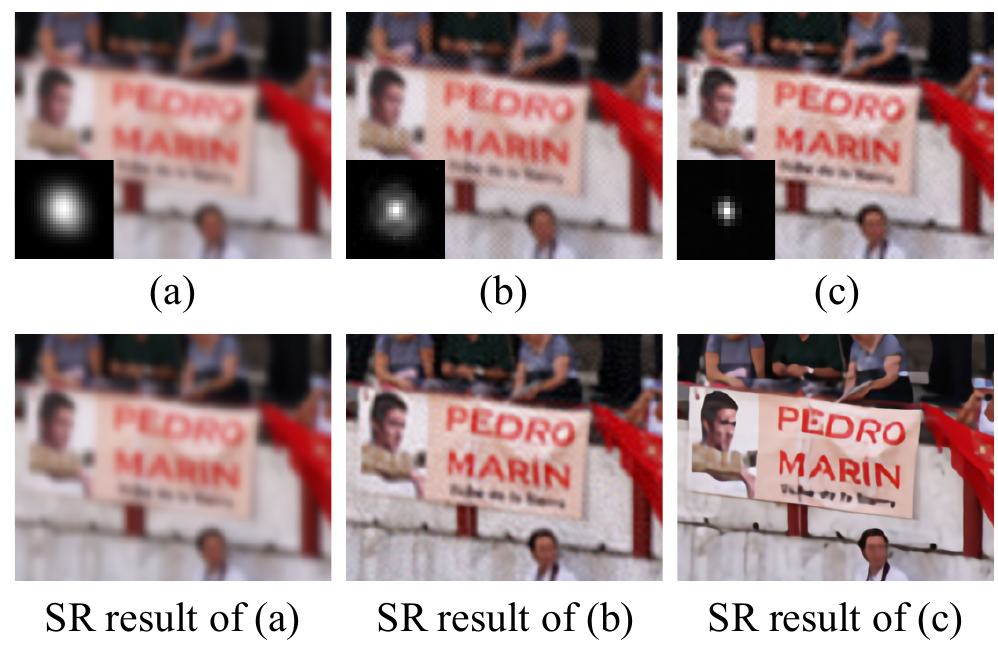}
\end{center}
\caption{Applying DCLS in the RGB space. (a) Original LR \& kernel, (b) corrected LR \& estimated kernel by~\cite{hussein2020correction}, (c) deblurred LR \& estimated kernel by the proposed method.}
\label{fig:dcls_rgb}
\end{figure}

\noindent \textbf{Effectiveness of the DCLS deconvolution.} 
To illustrate the effectiveness of DCLS, we include a comparison of substituting DCLS with other deblurring methods, such as traditional constrained least squares (CLS) and Wiener deconvolution~\cite{wiener1964extrapolation,dong2020deep} in the RGB space and feature space. The results are presented in Table~\ref{table:fea-deconv} and Table~\ref{table:rgb-deconv}. By applying deconvolution in the RGB space with the reformulated kernel, we can get a clear LR image and thus improve the SR performance. This idea is similar to Correction Filter~\cite{hussein2020correction}, but with one big difference, in that our estimator is highly correlated to the LR image rather than the SR model. The visual example is shown in Fig.~\ref{fig:dcls_rgb}.

\begin{figure}[t]
\setlength{\abovecaptionskip}{-0.1in}
\setlength{\belowcaptionskip}{-0.1in}
\begin{center}
\includegraphics[width=1.\linewidth]{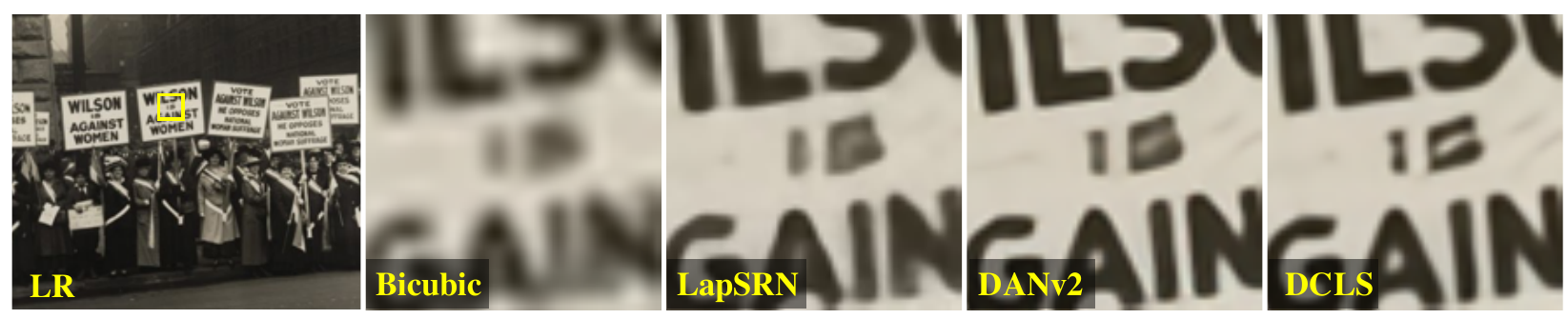}
\end{center}
\caption{Comparison of historic image~\cite{lai2017deep} for 4$\times$ SR.}
\label{fig:realim}
\end{figure}

\noindent \textbf{Performance on Real Degradation} 
To further demonstrate the effectiveness of our method, we apply the proposed model on real degradation data where the ground truth HR images and the blur kernels are not available. An example of super-resolving historic image is shown in Fig.~\ref{fig:realim}. Compared with LapSRN~\cite{lai2017deep} and DANv2~\cite{luo2021endtoend}, our DCLS can produce sharper edges and visual pleasing SR results.

\section{Conclusion}
In this work, we have presented a well-principled algorithm to tackle the blind SR problem. We first derive a new form of blur kernel in the low resolution space from classical degradation model. We then propose to estimate and apply that kernel in HR image restoration. Subsequently, a dynamic deep linear kernel (DDLK) module is introduced to improve kernel estimation. We further design a deep constrained least squares (DCLS) deconvolution module that integrates blur kernel and LR image in the feature domain to obtain the clean feature. The clean feature and the primitive feature are then fed into a dual-path network to generate the super-resolved image. Extensive experiments on various kernels and noises demonstrate that the proposed method leads to a state-of-the-art blind SR performance.

\noindent\textbf{Acknowledgment} This work was supported by the National Natural Science Foundation of China (NSFC) under grants
No.61872067 and No.61720106004.

{\small
\bibliographystyle{ieee_fullname}
\bibliography{egbib}

\begin{thebibliography}{10}\itemsep=-1pt

\bibitem{agustsson2017ntire}
Eirikur Agustsson and Radu Timofte.
\newblock Ntire 2017 challenge on single image super-resolution: Dataset and
  study.
\newblock In {\em {Proc. CVPRW}}, pages 126--135, 2017.

\bibitem{ahn2018fast}
Namhyuk Ahn, Byungkon Kang, and Kyung-Ah Sohn.
\newblock Fast, accurate, and lightweight super-resolution with cascading
  residual network.
\newblock In {\em {Proc. ECCV}}, pages 252--268, 2018.

\bibitem{bell2019blind}
Sefi Bell-Kligler, Assaf Shocher, and Michal Irani.
\newblock Blind super-resolution kernel estimation using an internal-gan.
\newblock In {\em {Proc. NeurIPS}}, pages 284--293, 2019.

\bibitem{bevilacqua2012low}
Marco Bevilacqua, Aline Roumy, Christine Guillemot, and Marie line
  Alberi~Morel.
\newblock Low-complexity single-image super-resolution based on nonnegative
  neighbor embedding.
\newblock In {\em {Proc. BMVC}}, pages 135.1--135.10, 2012.

\bibitem{dai2019second}
Tao Dai, Jianrui Cai, Yongbing Zhang, Shu-Tao Xia, and Lei Zhang.
\newblock Second-order attention network for single image super-resolution.
\newblock In {\em Proceedings of the IEEE/CVF Conference on Computer Vision and
  Pattern Recognition}, pages 11065--11074, 2019.

\bibitem{dong2014learning}
Chao Dong, Chen~Change Loy, Kaiming He, and Xiaoou Tang.
\newblock Learning a deep convolutional network for image super-resolution.
\newblock In {\em European conference on computer vision}, pages 184--199.
  Springer, 2014.

\bibitem{dong2020deep}
Jiangxin Dong, Stefan Roth, and Bernt Schiele.
\newblock Deep wiener deconvolution: Wiener meets deep learning for image
  deblurring.
\newblock {\em {Proc. NeurIPS}}, 33:1048--1059, 2020.

\bibitem{fritsche2019frequency}
Manuel Fritsche, Shuhang Gu, and Radu Timofte.
\newblock Frequency separation for real-world super-resolution.
\newblock In {\em 2019 IEEE/CVF International Conference on Computer Vision
  Workshop (ICCVW)}, pages 3599--3608. IEEE, 2019.

\bibitem{gu2019blind}
Jinjin Gu, Hannan Lu, Wangmeng Zuo, and Chao Dong.
\newblock Blind super-resolution with iterative kernel correction.
\newblock In {\em {Proc. CVPR}}, pages 1604--1613, 2019.

\bibitem{haris2018deep}
Muhammad Haris, Gregory Shakhnarovich, and Norimichi Ukita.
\newblock Deep back-projection networks for super-resolution.
\newblock In {\em Proceedings of the IEEE conference on computer vision and
  pattern recognition}, pages 1664--1673, 2018.

\bibitem{huang2009multi}
CK Huang and Hsiau-Hsian Nien.
\newblock Multi chaotic systems based pixel shuffle for image encryption.
\newblock {\em Optics communications}, 282(11):2123--2127, 2009.

\bibitem{huang2015single}
Jia-Bin Huang, Abhishek Singh, and Narendra Ahuja.
\newblock Single image super-resolution from transformed self-exemplars.
\newblock In {\em {Proc. CVPR}}, pages 5197--5206, 2015.

\bibitem{hussein2020correction}
Shady~Abu Hussein, Tom Tirer, and Raja Giryes.
\newblock Correction filter for single image super-resolution: Robustifying
  off-the-shelf deep super-resolvers.
\newblock In {\em Proceedings of the IEEE/CVF Conference on Computer Vision and
  Pattern Recognition}, pages 1428--1437, 2020.

\bibitem{jo2021adatarget}
Younghyun Jo, Seoung~Wug Oh, Peter Vajda, and Seon~Joo Kim.
\newblock Tackling the ill-posedness of super-resolution through adaptive
  target generation.
\newblock In {\em {Proc. CVPR}}, pages 16236--16245, 2021.

\bibitem{johnson2016perceptual}
Justin Johnson, Alexandre Alahi, and Li Fei-Fei.
\newblock Perceptual losses for real-time style transfer and super-resolution.
\newblock In {\em European conference on computer vision}, pages 694--711.
  Springer, 2016.

\bibitem{kawaguchi2016deep}
Kenji Kawaguchi.
\newblock Deep learning without poor local minima.
\newblock In {\em {Proc. NeurIPS}}, pages 586--594, 2016.

\bibitem{kim2016accurate}
Jiwon Kim, Jung~Kwon Lee, and Kyoung~Mu Lee.
\newblock Accurate image super-resolution using very deep convolutional
  networks.
\newblock In {\em Proceedings of the IEEE conference on computer vision and
  pattern recognition}, pages 1646--1654, 2016.

\bibitem{kim2016deeply}
Jiwon Kim, Jung~Kwon Lee, and Kyoung~Mu Lee.
\newblock Deeply-recursive convolutional network for image super-resolution.
\newblock In {\em Proceedings of the IEEE conference on computer vision and
  pattern recognition}, pages 1637--1645, 2016.

\bibitem{kim2021koalanet}
Soo~Ye Kim, Hyeonjun Sim, and Munchurl Kim.
\newblock Koalanet: Blind super-resolution using kernel-oriented adaptive local
  adjustment.
\newblock In {\em {Proc. CVPR}}, pages 10611--10620, 2021.

\bibitem{kingma2014adam}
Diederik~P Kingma and Jimmy Ba.
\newblock Adam: A method for stochastic optimization.
\newblock {\em arXiv preprint arXiv:1412.6980}, 2014.

\bibitem{lai2017deep}
Wei-Sheng Lai, Jia-Bin Huang, Narendra Ahuja, and Ming-Hsuan Yang.
\newblock Deep laplacian pyramid networks for fast and accurate
  super-resolution.
\newblock In {\em Proceedings of the IEEE conference on computer vision and
  pattern recognition}, pages 624--632, 2017.

\bibitem{ledig2017photo}
Christian Ledig, Lucas Theis, Ferenc Husz{\'a}r, Jose Caballero, Andrew
  Cunningham, Alejandro Acosta, Andrew Aitken, Alykhan Tejani, Johannes Totz,
  Zehan Wang, et~al.
\newblock Photo-realistic single image super-resolution using a generative
  adversarial network.
\newblock In {\em Proceedings of the IEEE conference on computer vision and
  pattern recognition}, pages 4681--4690, 2017.

\bibitem{lee2018towards}
Guang-He Lee, David Alvarez-Melis, and Tommi~S Jaakkola.
\newblock Towards robust, locally linear deep networks.
\newblock In {\em {Proc. ICLR}}, 2018.

\bibitem{levin2009understanding}
Anat Levin, Yair Weiss, Fredo Durand, and William~T Freeman.
\newblock Understanding and evaluating blind deconvolution algorithms.
\newblock In {\em 2009 IEEE Conference on Computer Vision and Pattern
  Recognition}, pages 1964--1971. IEEE, 2009.

\bibitem{levin2011efficient}
Anat Levin, Yair Weiss, Fredo Durand, and William~T Freeman.
\newblock Efficient marginal likelihood optimization in blind deconvolution.
\newblock In {\em CVPR 2011}, pages 2657--2664. IEEE, 2011.

\bibitem{li2019feedback}
Zhen Li, Jinglei Yang, Zheng Liu, Xiaomin Yang, Gwanggil Jeon, and Wei Wu.
\newblock Feedback network for image super-resolution.
\newblock In {\em Proceedings of the IEEE/CVF Conference on Computer Vision and
  Pattern Recognition}, pages 3867--3876, 2019.

\bibitem{liang2021flow}
Jingyun Liang, Kai Zhang, Shuhang Gu, Luc Van~Gool, and Radu Timofte.
\newblock Flow-based kernel prior with application to blind super-resolution.
\newblock In {\em Proceedings of the IEEE/CVF Conference on Computer Vision and
  Pattern Recognition}, pages 10601--10610, 2021.

\bibitem{lim2017enhanced}
Bee Lim, Sanghyun Son, Heewon Kim, Seungjun Nah, and Kyoung Mu~Lee.
\newblock Enhanced deep residual networks for single image super-resolution.
\newblock In {\em {Proc. CVPRW}}, pages 136--144, 2017.

\bibitem{lugmayr2020srflow}
Andreas Lugmayr, Martin Danelljan, Luc Van~Gool, and Radu Timofte.
\newblock Srflow: Learning the super-resolution space with normalizing flow.
\newblock In {\em European Conference on Computer Vision}, pages 715--732.
  Springer, 2020.

\bibitem{luo2020unfolding}
Zhengxiong Luo, Yan Huang, Shang Li, Liang Wang, and Tieniu Tan.
\newblock Unfolding the alternating optimization for blind super resolution.
\newblock In {\em {Proc. NeurIPS}}, 2020.

\bibitem{luo2021endtoend}
Zhengxiong Luo, Yan Huang, Shang Li, Liang Wang, and Tieniu Tan.
\newblock End-to-end alternating optimization for blind super resolution.
\newblock {\em arXiv preprint arXiv:2105.06878}, 2021.

\bibitem{luo2021ebsr}
Ziwei Luo, Lei Yu, Xuan Mo, Youwei Li, Lanpeng Jia, Haoqiang Fan, Jian Sun, and
  Shuaicheng Liu.
\newblock Ebsr: Feature enhanced burst super-resolution with deformable
  alignment.
\newblock In {\em Proceedings of the IEEE/CVF Conference on Computer Vision and
  Pattern Recognition}, pages 471--478, 2021.

\bibitem{martin2001database}
David Martin, Charless Fowlkes, Doron Tal, and Jitendra Malik.
\newblock A database of human segmented natural images and its application to
  evaluating segmentation algorithms and measuring ecological statistics.
\newblock In {\em {Proc. ICCV}}, pages 416--423, 2001.

\bibitem{matsui2017sketch}
Yusuke Matsui, Kota Ito, Yuji Aramaki, Azuma Fujimoto, Toru Ogawa, Toshihiko
  Yamasaki, and Kiyoharu Aizawa.
\newblock Sketch-based manga retrieval using manga109 dataset.
\newblock {\em Multimedia Tools and Applications}, 76(20):21811--21838, 2017.

\bibitem{michaeli2013nonparametric}
Tomer Michaeli and Michal Irani.
\newblock Nonparametric blind super-resolution.
\newblock In {\em Proceedings of the IEEE International Conference on Computer
  Vision}, pages 945--952, 2013.

\bibitem{montufar2014number}
Guido Mont{\'u}far, Razvan Pascanu, Kyunghyun Cho, and Yoshua Bengio.
\newblock On the number of linear regions of deep neural networks.
\newblock In {\em {Proc. NeurIPS}}, pages 2924--2932, 2014.

\bibitem{pan2017deblurring}
Jinshan Pan, Deqing Sun, Hanspeter Pfister, and Ming-Hsuan Yang.
\newblock Deblurring images via dark channel prior.
\newblock {\em {IEEE Trans. on Pattern Analysis and Machine Intelligence}},
  40(10):2315--2328, 2017.

\bibitem{park2003super}
Sung~Cheol Park, Min~Kyu Park, and Moon~Gi Kang.
\newblock Super-resolution image reconstruction: a technical overview.
\newblock {\em IEEE signal processing magazine}, 20(3):21--36, 2003.

\bibitem{saxe2013exact}
Andrew~M Saxe, James~L McClelland, and Surya Ganguli.
\newblock Exact solutions to the nonlinear dynamics of learning in deep linear
  neural networks.
\newblock {\em arXiv preprint arXiv:1312.6120}, 2013.

\bibitem{shi2016real}
Wenzhe Shi, Jose Caballero, Ferenc Husz{\'a}r, Johannes Totz, Andrew~P Aitken,
  Rob Bishop, Daniel Rueckert, and Zehan Wang.
\newblock Real-time single image and video super-resolution using an efficient
  sub-pixel convolutional neural network.
\newblock In {\em Proceedings of the IEEE conference on computer vision and
  pattern recognition}, pages 1874--1883, 2016.

\bibitem{shocher2018zero}
Assaf Shocher, Nadav Cohen, and Michal Irani.
\newblock “zero-shot” super-resolution using deep internal learning.
\newblock In {\em {Proc. CVPR}}, pages 3118--3126, 2018.

\bibitem{soh2020meta}
Jae~Woong Soh, Sunwoo Cho, and Nam~Ik Cho.
\newblock Meta-transfer learning for zero-shot super-resolution.
\newblock In {\em Proceedings of the IEEE/CVF Conference on Computer Vision and
  Pattern Recognition}, pages 3516--3525, 2020.

\bibitem{tai2017image}
Ying Tai, Jian Yang, and Xiaoming Liu.
\newblock Image super-resolution via deep recursive residual network.
\newblock In {\em Proceedings of the IEEE conference on computer vision and
  pattern recognition}, pages 3147--3155, 2017.

\bibitem{tao2021spectrum}
Guangpin Tao, Xiaozhong Ji, Wenzhuo Wang, Shuo Chen, Chuming Lin, Yun Cao, Tong
  Lu, Donghao Luo, and Ying Tai.
\newblock Spectrum-to-kernel translation for accurate blind image
  super-resolution.
\newblock In {\em Thirty-Fifth Conference on Neural Information Processing
  Systems}, 2021.

\bibitem{timofte2017ntire}
Radu Timofte, Eirikur Agustsson, Luc Van~Gool, Ming-Hsuan Yang, and Lei Zhang.
\newblock Ntire 2017 challenge on single image super-resolution: Methods and
  results.
\newblock In {\em {Proc. CVPRW}}, pages 114--125, 2017.

\bibitem{vandewalle2006frequency}
Patrick Vandewalle, Sabine S{\"u}sstrunk, and Martin Vetterli.
\newblock A frequency domain approach to registration of aliased images with
  application to super-resolution.
\newblock {\em EURASIP journal on advances in signal processing}, 2006:1--14,
  2006.

\bibitem{wang2018recovering}
Xintao Wang, Ke Yu, Chao Dong, and Chen~Change Loy.
\newblock Recovering realistic texture in image super-resolution by deep
  spatial feature transform.
\newblock In {\em Proceedings of the IEEE conference on computer vision and
  pattern recognition}, pages 606--615, 2018.

\bibitem{wang2018esrgan}
Xintao Wang, Ke Yu, Shixiang Wu, Jinjin Gu, Yihao Liu, Chao Dong, Yu Qiao, and
  Chen Change~Loy.
\newblock Esrgan: Enhanced super-resolution generative adversarial networks.
\newblock In {\em Proceedings of the European conference on computer vision
  (ECCV) workshops}, pages 0--0, 2018.

\bibitem{wang2004image}
Zhou Wang, Alan~C Bovik, Hamid~R Sheikh, and Eero~P Simoncelli.
\newblock Image quality assessment: from error visibility to structural
  similarity.
\newblock {\em {IEEE Trans. on Image Processing}}, 13(4):600--612, 2004.

\bibitem{wiener1964extrapolation}
Norbert Wiener et~al.
\newblock {\em Extrapolation, interpolation, and smoothing of stationary time
  series: with engineering applications}, volume~8.
\newblock MIT press Cambridge, MA, 1964.

\bibitem{xu2020unified}
Yu-Syuan Xu, Shou-Yao~Roy Tseng, Yu Tseng, Hsien-Kai Kuo, and Yi-Min Tsai.
\newblock Unified dynamic convolutional network for super-resolution with
  variational degradations.
\newblock In {\em Proceedings of the IEEE/CVF Conference on Computer Vision and
  Pattern Recognition}, pages 12496--12505, 2020.

\bibitem{yu2016ultra}
Xin Yu and Fatih Porikli.
\newblock Ultra-resolving face images by discriminative generative networks.
\newblock In {\em European conference on computer vision}, pages 318--333.
  Springer, 2016.

\bibitem{zeyde2010single}
Roman Zeyde, Michael Elad, and Matan Protter.
\newblock On single image scale-up using sparse-representations.
\newblock In {\em International Conference on Curves and Surfaces}, pages
  711--730, 2010.

\bibitem{zhang2020deep}
Kai Zhang, Luc~Van Gool, and Radu Timofte.
\newblock Deep unfolding network for image super-resolution.
\newblock In {\em {Proc. CVPR}}, pages 3217--3226, 2020.

\bibitem{zhang2017beyond}
Kai Zhang, Wangmeng Zuo, Yunjin Chen, Deyu Meng, and Lei Zhang.
\newblock Beyond a gaussian denoiser: Residual learning of deep cnn for image
  denoising.
\newblock {\em {IEEE Trans. on Image Processing}}, 26(7):3142--3155, 2017.

\bibitem{zhang2018learning}
Kai Zhang, Wangmeng Zuo, and Lei Zhang.
\newblock Learning a single convolutional super-resolution network for multiple
  degradations.
\newblock In {\em {Proc. CVPR}}, pages 3262--3271, 2018.

\bibitem{zhang2019deep}
Kai Zhang, Wangmeng Zuo, and Lei Zhang.
\newblock Deep plug-and-play super-resolution for arbitrary blur kernels.
\newblock In {\em {Proc. CVPR}}, pages 1671--1681, 2019.

\bibitem{zhang2018unreasonable}
Richard Zhang, Phillip Isola, Alexei~A Efros, Eli Shechtman, and Oliver Wang.
\newblock The unreasonable effectiveness of deep features as a perceptual
  metric.
\newblock In {\em Proceedings of the IEEE conference on computer vision and
  pattern recognition}, pages 586--595, 2018.

\bibitem{zhang2018image}
Yulun Zhang, Kunpeng Li, Kai Li, Lichen Wang, Bineng Zhong, and Yun Fu.
\newblock Image super-resolution using very deep residual channel attention
  networks.
\newblock In {\em {Proc. ECCV}}, pages 286--301, 2018.

\bibitem{Zhang_2018_ECCV}
Yulun Zhang, Kunpeng Li, Kai Li, Lichen Wang, Bineng Zhong, and Yun Fu.
\newblock Image super-resolution using very deep residual channel attention
  networks.
\newblock In {\em {Proc. ECCV}}, pages 294--310, 2018.

\bibitem{zhang2018residual}
Yulun Zhang, Yapeng Tian, Yu Kong, Bineng Zhong, and Yun Fu.
\newblock Residual dense network for image super-resolution.
\newblock In {\em Proceedings of the IEEE conference on computer vision and
  pattern recognition}, pages 2472--2481, 2018.

\bibitem{zhang2020residual}
Yulun Zhang, Yapeng Tian, Yu Kong, Bineng Zhong, and Yun Fu.
\newblock Residual dense network for image restoration.
\newblock {\em IEEE Transactions on Pattern Analysis and Machine Intelligence},
  43(7):2480--2495, 2020.

\end{thebibliography}
}

\end{document}